\documentclass[sigconf, nonacm]{acmart}

\usepackage{amsmath,amsfonts}
\usepackage{algorithm}
\usepackage{balance}
\usepackage{algorithmic}
\usepackage{graphicx}
\usepackage{listings}
\usepackage{mathrsfs}
\usepackage{mdwlist}
\usepackage{textcomp}
\usepackage{xcolor}
\usepackage{xspace}

\usepackage{subcaption}

\usepackage[nolist,nohyperlinks]{acronym}
\begin{acronym}
\acro{DB}{database}
\end{acronym}

\usepackage{xspace}
\newcommand{\PaperAcronym}{DiffML\xspace}

\begin{document}
\title{\PaperAcronym: End-to-end Differentiable ML Pipelines}

\author{Benjamin Hilprecht$^*$}
\affiliation{\institution{TU Darmstadt}
}

\author{Christian Hammacher$^*$}
\affiliation{\institution{Software AG}
}

\author{Eduardo Reis}
\affiliation{\institution{TU Darmstadt}
}

\author{Mohamed Abdelaal}
\affiliation{\institution{Software AG}
}

\author{Carsten Binnig}
\affiliation{\institution{TU Darmstadt}
}

\begin{abstract}

In this paper, we present our vision of differentiable ML pipelines called \PaperAcronym{} to automate the construction of ML pipelines in an end-to-end fashion. 
The idea is that \PaperAcronym{} allows to jointly train not just the ML model itself but also the entire pipeline including data preprocessing steps, e.g., data cleaning, feature selection, etc. Our core idea is to formulate all pipeline steps in a differentiable way such that the entire pipeline can be trained using backpropagation. However, this is a non-trivial problem and opens up many new research questions.
To show the feasibility of this direction, we demonstrate initial ideas and a general principle of how typical preprocessing steps such as data cleaning, feature selection and dataset selection can be formulated as differentiable programs and jointly learned with the ML model. Moreover, we discuss a research roadmap and core challenges that have to be systematically tackled to enable fully differentiable ML pipelines.

\end{abstract}

\maketitle

\def\thefootnote{*}\footnotetext{These authors contributed equally to this work}\def\thefootnote{\arabic{footnote}}

\begin{figure}
	\centering
	\includegraphics[width=0.8\columnwidth]{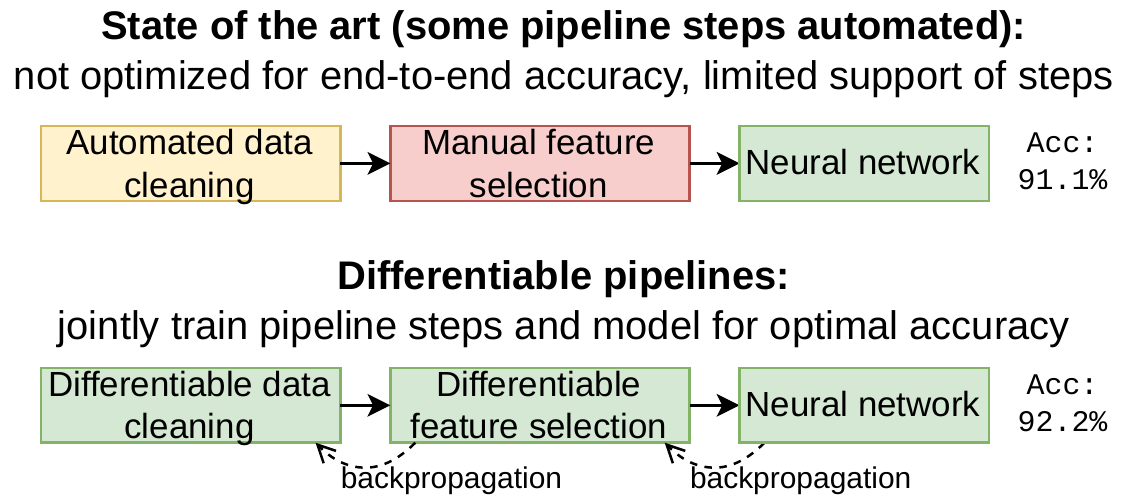}
	\vspace{-2.5ex}
	\caption{Overview of differentiable ML pipelines. Traditionally, only some pipeline steps could be derived automatically and these typically do not directly target the end-to-end performance. Differentiable pipelines instead allow training all steps in the pipeline (data cleaning, feature selection, etc.) jointly with the ML model for an improved overall accuracy.}
	\vspace{-3.5ex}
	\label{fig:overview}
\end{figure}

\section{Introduction}

In recent years, there have been many breakthroughs in machine learning (ML).
However, developing ML pipelines to solve particular tasks is still far from trivial and requires expertise in many areas. 
In particular, besides training a model, to develop an ML pipeline, we not only have to tackle core tasks regarding model selection and hyperparameter tuning, but we also have to solve many tasks related to data engineering such as data cleaning, preprocessing, feature selection etc. These tasks typically require a high-level of expertise and are thus a barrier for the broad adoption of ML. Furthermore, even with expert knowledge, these steps are highly time-consuming since they require many manual decisions and experimentation.

Hence, recent efforts concentrate on automating the steps of typical ML pipelines and in particular also data engineering.
For instance, it was proposed to automate data cleaning using ML techniques in particular to detect \cite{10.14778/2994509.2994518,DBLP:conf/sigmod/MahdaviAFMOS019,DBLP:conf/cikm/NeutatzMA19,dboost} or correct data errors \cite{10.14778/3137628.3137631,10.14778/3407790.3407801}, impute missing values \cite{10.14778/3137628.3137631,mlsys2020_123,pmlr-v80-yoon18a}, or even compensate a potential bias in the data \cite{DBLP:conf/sigmod/OrrBS20,DBLP:conf/sigmod/HilprechtB21}.

However, while these automation techniques typically help with the individual tasks, it is not guaranteed that they improve the end-to-end performance of the ML pipeline. For instance, different papers have shown that while automated data cleaning can improve the performance, an unsuitable cleaning method can also significantly deteriorate the accuracy of a (downstream) model \cite{neutatz2021cleaning,DBLP:conf/icde/LiRBZCZ21,neutatz2022data}. The reason behind this deterioration is that the automated individual techniques often do not target the final accuracy of the ML model but metrics of the particular task (e.g., the accuracy of the imputation strategy). 

\vspace{-1.5ex}\paragraph{Vision and Contributions}
In this paper, we thus propose our vision of \PaperAcronym, where our goal is to automatically construct the \textit{entire} ML pipeline holistically while optimizing the \textit{end-to-end performance}. The key idea is to train the ML pipeline steps end-to-end and jointly with the actual model, which ensures that the ML pipeline is suited for the task at hand and optimized for accuracy (cf. Figure~\ref{fig:overview}). In particular, to enable trainable ML pipelines, we propose to express those pipelines as a differentiable program.
That way, we can train all steps of an ML pipeline end-to-end; e.g., we can train the data cleaning steps jointly with the ML model by using backpropagation along the full pipeline.

However, it is not straightforward to formulate entire ML pipelines as a differentiable program.
As a concrete contribution in this paper, we discuss a concrete direction of using so-called \emph{mixtures of pipeline alternatives} (cf. Section \ref{sec:overview}) to enable differentiable ML pipelines. 
To show the generality of using the idea of mixtures for enabling differentiable ML pipelines, we demonstrate how various pipeline steps can be trained jointly with an ML model by making them differentiable using this idea.
As concrete use cases to show the feasibility of our approach, we implemented three concrete scenarios which differ in the nature of the preprocessing steps that are combined with the model training. 
Furthermore, we think that this paper is only a starting point and more additional research is required to automate other pipeline steps or even explore other directions to enable differentiable ML pipelines beyond the idea of using mixtures.

Note that there have been previous attempts to derive ML pipelines holistically where however important pipeline steps are usually not integrated. For instance, AutoML \cite{feurer2015efficient,shang2019democratizing,DBLP:journals/jair/ZollerH21} typically only automates model selection and hyperparameter tuning, whereas for instance data cleaning is typically not considered. Moreover, a major advantage of our approach is that just a single pipeline has to be trained whereas AutoML \cite{DBLP:journals/jair/ZollerH21} requires searching the space of potential ML pipelines by training multiple pipelines (e.g., with different data preprocessing steps, different model architectures, and different hyperparameters), which increases the computational overhead significantly. Similarly, it was recently proposed to fuse several ML pipeline steps into a single differentiable program \cite{10.14778/3485450.3485452}.
However, in contrast to \PaperAcronym, more complex pipeline steps such as data cleaning are again not supported in \cite{10.14778/3485450.3485452}, which are often crucial for a competitive end-to-end performance. 

\vspace{-1.5ex}\paragraph{Outline}
In Section~\ref{sec:overview}, we present our vision of expressing the various steps in ML pipelines as a single differentiable program and discuss the advantages. Afterwards, in Section~\ref{sec:case_studies}, we present more details on how the general framework of \PaperAcronym{} can be applied to the problems of data cleaning, dataset and feature selection and show promising initial results that suggest that differentiable ML pipelines can achieve competitive ML performance with only a single training phase. In Section~\ref{sec:res_challenges}, we present a research roadmap to enable differentiable ML pipelines for a broad set of tasks and finally provide an outlook in Section~\ref{sec:concl}. \section{Differentiable ML Pipelines}
\label{sec:overview}

In the following, we give an overview of \PaperAcronym{} and discuss both the advantages of such a design and the key challenges to realize this vision.

\subsection{Overview of the Approach}
\label{sec:overview_approach}

\begin{figure}
	\centering
	\includegraphics[width=0.9\columnwidth]{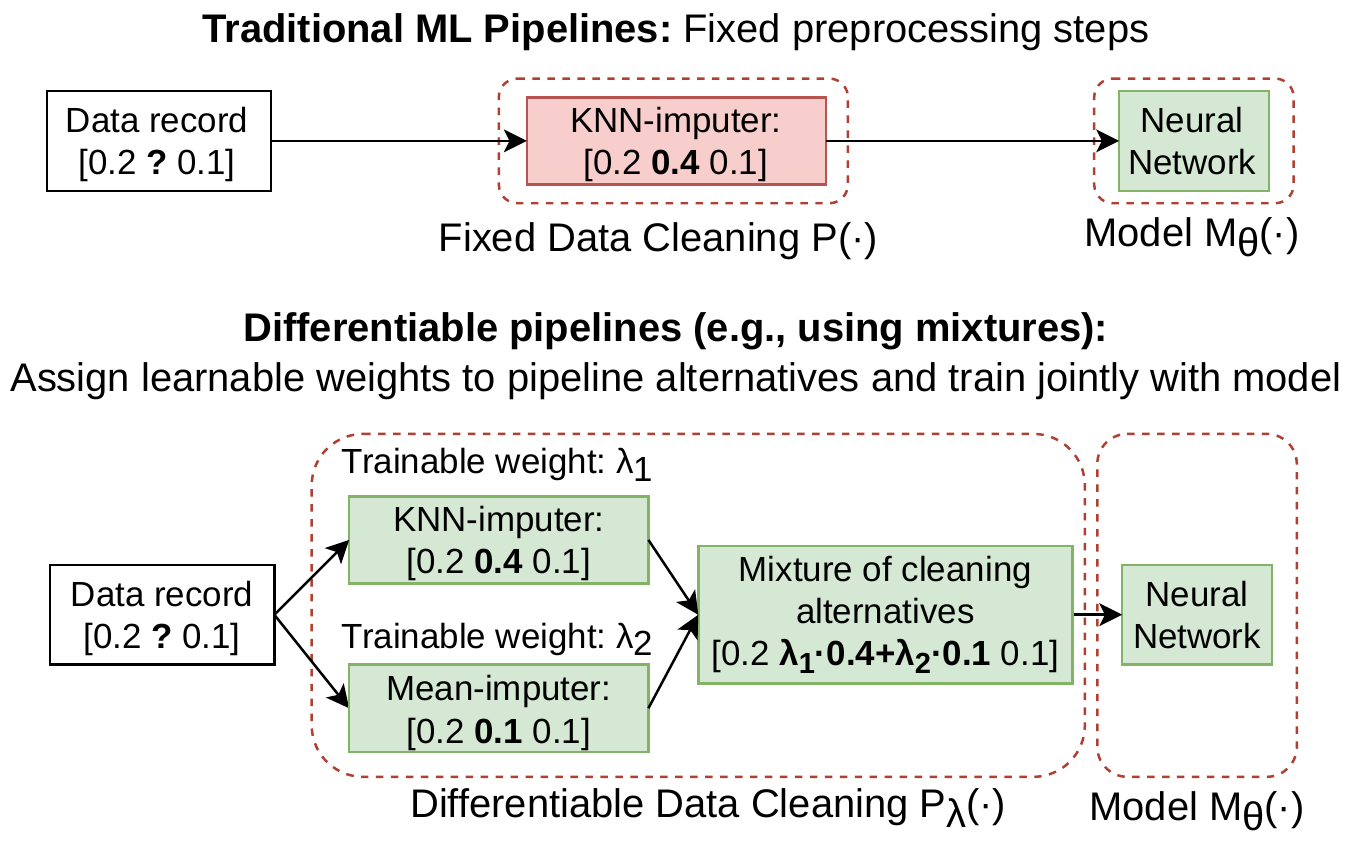}
	\vspace{-3.5ex}
	\caption{Differentiable pipelines can be realized as learning mixtures of alternatives for pipeline steps. This way, we can learn which preprocessing steps are suitable to improve the end-to-end model performance. For example, we could learn alternatives such as $\lambda_1=1,\lambda_2=0$ which means that only the KNN-imputer is used whereas $\lambda_1=0,\lambda_2=1$ uses only the mean imputer. However, we can also learn to combine both imputers (e.g., for $\lambda_1=\lambda_2=0.5$).}
	\vspace{-5.5ex}
	\label{fig:example}
\end{figure}

Figure \ref{fig:example} shows the high-level idea of \PaperAcronym{}: we parameterize the search space of preprocessing steps in a pipeline and train it jointly with the ML model itself. This way, the preprocessing pipeline best suited for the model and the particular task is instantiated. 

A major challenge to enable the vision of \PaperAcronym{} is that many steps of ML pipelines can not be trivially expressed in a differentiable way.
The reasons behind this challenge are twofold:
First, for each pipeline step, users can typically choose from very different methods. For example, various methods exist to impute missing values, e.g., mean imputation, KNN-based imputation, etc.
Overall, selecting the best method from a set of methods is a non-differentiable problem by nature.
Second, as mentioned before, for some individual pipeline steps, there already exist learned methods which themselves are differentiable (e.g., learned data cleaning). 
However, integrating these learned approaches into a differentiable ML pipeline and training them end-to-end with an ML model is not trivial, since the optimization objectives typically differ.  

Hence, the main idea of \PaperAcronym{} to make ML pipelines differentiable is to express different alternatives as a mixture of pipeline alternatives, where during the training process the weights for the different alternatives are learned (cf. Figure~\ref{fig:example}) (lower part). 
Moreover, we believe that mixtures are just one way and many more options exist, as we discuss in Section \ref{sec:res_challenges}.
In the following, to illustrate the idea, we first sketch how ML pipelines are trained today before we introduce our approach based on the idea of mixtures to achieve differentiable ML pipelines.

\vspace{-1.5ex}\paragraph{ML Pipelines today} Traditionally, the preprocessing steps $P$ in an ML pipeline are typically considered fixed and only the ML model $M_\theta$ itself is trained. Hence, to obtain a prediction for a particular example $x$ in our dataset, we first apply the preprocessing steps and feed the result into our model: $M_\theta(P(x)).$ During training, we find the parameters $\theta$ that minimize our loss. For instance, the preprocessing could just be a missing value imputation, where missing values are replaced by some constant dummy values. If we now want to evaluate a different preprocessing pipeline $P'$ such as K-nearest neighbors (KNN) for replacing missing values, we would train a different model $M_{\theta'}$ with different parameters $\theta'$ and compare the performance of both pipelines on the validation set. Such retraining can be costly, since the search space of different pipelines grows exponentially as more steps are considered.

\vspace{-1.5ex}\paragraph{ML Pipelines with \PaperAcronym{}} Instead in our approach, we not only parameterize the model $M_\theta$ but also the preprocessing step $P_\lambda(x)$ and optimize both sets of parameters to also train the preprocessing step itself. As mentioned before, we require $P_\lambda(x)$ to be differentiable w.r.t. $\lambda$ s.t. a gradient-descent-based optimization can be applied. Unfortunately, expressing pipelines steps as differentiable programs is often non-trivial. However, we believe that often different options can be combined using mixtures, where during training the weights of different alternatives can be learned.

For instance in Figure \ref{fig:example}, instead of using a fixed procedure to impute missing values, we use a mixture of KNN-based and mean imputation and learn the weights $\lambda$ of each approach, i.e., we feed the weighted sum of both imputations into the model. In the extreme case of using $\lambda_0=0$ and $\lambda_1=1$, this reduces to just using a fixed mean imputation in our pipeline. As such, the case of using just a single imputation strategy is simply a special case of this formulation and can still be expressed. In addition, this approach also enables combinations of different imputation strategies (e.g., by using $\lambda_0=\lambda_1=0.5$) which can be beneficial, as we will show in our initial case studies.

\subsection{Discussion}

As discussed before, our goal is to train the entire ML pipeline using gradient-based optimization, which basically requires that the ML model itself is also trainable using methods such SGD or similar techniques. While this is true for many popular classes of models including DNNs, linear and logistic regression or support vector machines (SVMs), there are other model types which do not directly adhere to this training regime such as decision trees. 

However, recently it was shown that a much broader class of models can be made differentiable by translating the models (e.g., decision trees) into NN layers \cite{10.14778/3485450.3485452}. This technique can also be applied directly in our setup to enable a more diverse set of models. Moreover, motivated by the breakthroughs achieved using DNNs, we believe that it is already very attractive to only focus on DNNs in an initial realization of our vision, which we will be our focus in the remainder of this paper.

Besides having a differentiable model, a key challenge to enable our vision is to express the different steps of ML pipelines in a differentiable way. Hence, to demonstrate the feasibility we will next showcase how three pipeline steps can in fact be formulated in a differentiable manner, namely data cleaning, dataset selection and feature selection. We will discuss the challenges of additional operators as well as other potential future directions in Section~\ref{sec:res_challenges}. 

 \section{Case Studies}
\label{sec:case_studies}

In the following, we present how three important preprocessing steps (data cleaning, dataset selection and feature selection)  can be trained jointly with the ML model by expressing them as differentiable programs.

\begin{figure}
	\centering
	\includegraphics[width=0.8\columnwidth]{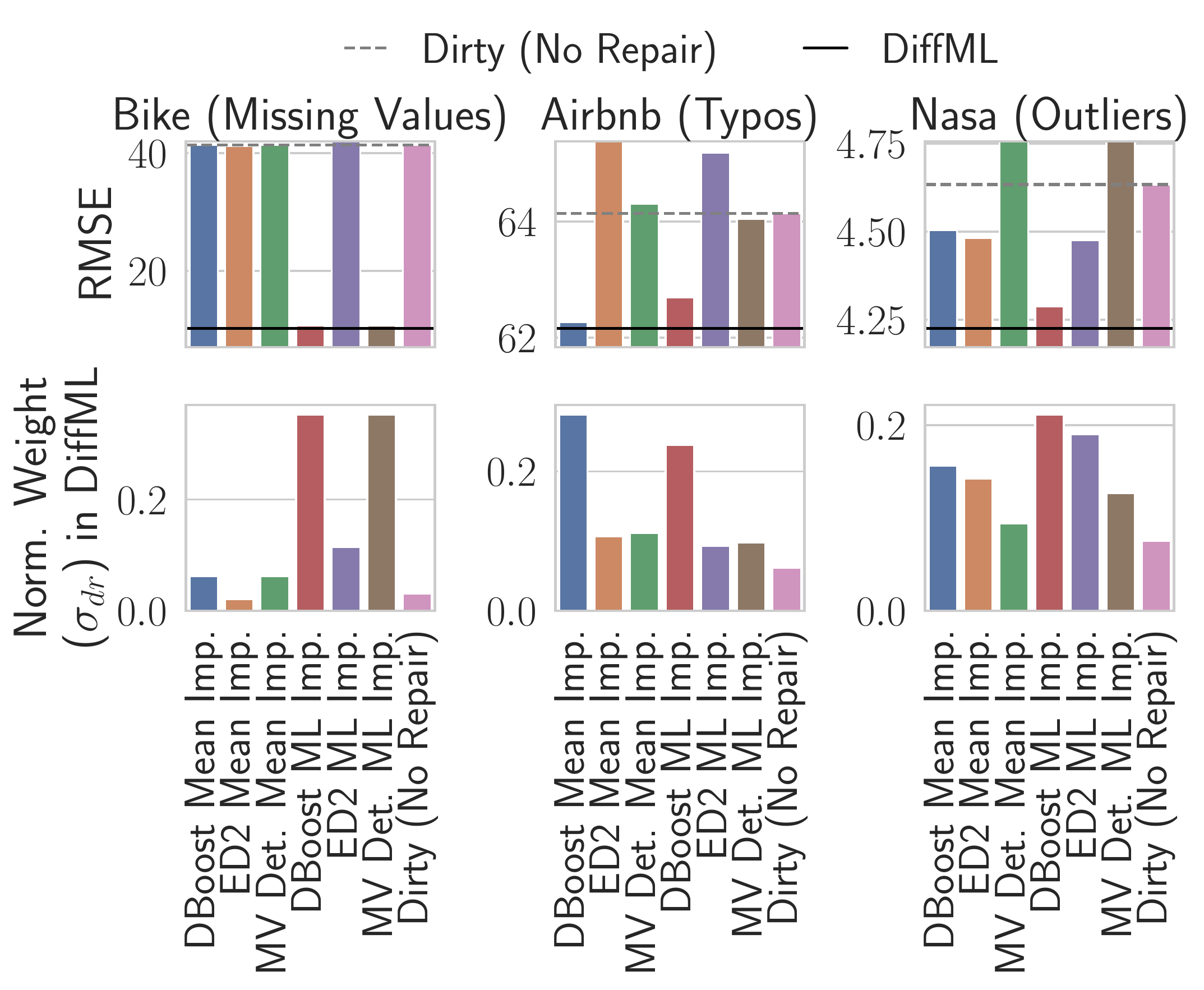}
	\vspace{-3.5ex}
	\caption{Prediction errors on dirty datasets for data cleaning methods and \PaperAcronym (black line, top) and learned weights for data cleaning methods (bottom) in \PaperAcronym. \PaperAcronym assigns higher weights to more effective data cleaning methods and thus always achieves a competitive performance. Moreover, for finding these weights, \PaperAcronym only requires to train a single pipeline. }
	\vspace{-2.5ex}
	\label{fig:exp_data_cleaning}
\end{figure}

\subsection{Learned Data Cleaning}
\label{sec:learned_data_cleaning}

In this section, we aim to learn which data cleaning procedure to use jointly with the actual model training. Before we explain how our approach works, we provide necessary background.

\vspace{-1.5ex}\paragraph{Background.} Data cleaning \cite{chu2016data} is usually organized in two subtasks: error detection \cite{10.14778/2994509.2994518,DBLP:conf/sigmod/MahdaviAFMOS019,DBLP:conf/cikm/NeutatzMA19,dboost} and repair \cite{10.14778/3137628.3137631,10.14778/3407790.3407801}. In the first task, we aim to identify which values in a tuple are erroneous. Having identified the errors, a repair method then aims to predict the correct values before replacing such values in the dataset. Hence, for a combination of a detection method $d$ and a repair method $r$, we obtain a different repaired tuple $x_{dr}$ for every original tuple $x$ in our dataset. Traditionally, to find an effective detection and repair method, different ML pipelines with different data cleaning methods had to be trained. 

\vspace{-1.5ex}\paragraph{Method.} We now show how to express data cleaning as a differentiable program based on the general idea of mixtures introduced in Section~\ref{sec:overview_approach}. More precisely, we formulate the search space of different error detection and repair methods as a mixture where every detection and repair method is assigned a weight $\lambda_d$ and $\lambda_r,$ respectively. A particular tuple $x_{dr}$ should then be weighted by $\lambda_d+\lambda_r$. During training, we will assign higher weights to more effective pairs. The weighted sum of all repaired tuples $\sum_{d\in D,r\in R} (\lambda_d+\lambda_r)x_{dr}$ could now simply be fed into the model to make the entire pipeline differentiable. 

To approximate a probability distribution for the learned weights, we do not directly use the weights $\lambda_d+\lambda_r$ but the softmax, i.e., 
$\sigma_{dr} = \frac{\mathrm{e}^{(\lambda_d+\lambda_r)}}{\sum_{d'\in D,r'\in R}\mathrm{e}^{\lambda_{d'}+\lambda_{r'}}}$.
For learning and inference, we use a tuple $x_{dr}$ with renormalized weights as an input to the model $M$: $M_\theta\left(\sum_{d\in D,r\in R} \sigma_{dr}x_{dr}\right)$. During training, we can thus jointly learn the model parameters $\theta$ and weights $\lambda$ by computing a loss function on the output and then computing gradients for both during backpropagation.
We also found that it is also slightly beneficial to not learn both the weights $\lambda$ and the model parameters $\theta$ in a single pass, but in each step of mini-batch learning sample two different batches and update the parameters sequentially.

\vspace{-1.5ex}\paragraph{Initial Results.} 
We study the effectiveness of our approach in an initial experiment where the task is to solve a regression problem on tabular data. In particular, we consider three datasets where each of them has a specific error type which is common in practice: \emph{Bike} with missing values, \emph{Nasa} with outliers and \emph{Airbnb} with typos on 10\% of the values each.

To evaluate how state-of-the-art data cleaning can improve the performance of the downstream model, we repaired the datasets using existing state-of-the-art detection methods, including DBoost \cite{dboost}, ED2 \cite{DBLP:conf/cikm/NeutatzMA19} and a simple missing value detector. To repair the errors, we used an ML-based imputation strategy (KNN) as well as an imputer using the mean. We then trained a Multilayer Perceptron (MLP) on the dirty data as well as the repaired versions and compare the root mean square error (RMSE). In addition, we evaluate \PaperAcronym{} which can choose from all of these detection and repair methods. As an additional baseline, we also train the model on the dirty version of the dataset, i.e., without any cleaning.\footnote{We did not compare to approaches such as AutoML or fused ML pipelines \cite{10.14778/3485450.3485452} since they only support simple preprocessing steps such as feature normalization but not complex steps as we support in \PaperAcronym{}.}

As we can see in Figure~\ref{fig:exp_data_cleaning} (top chart), \PaperAcronym{} achieves RMSE values (black solid line) which are competitive compared to the best method for each of the datasets. Importantly, \PaperAcronym{} trains only a single pipeline whereas traditionally multiple pipelines would have to be trained to identify the most effective data cleaning strategy.
Moreover, when comparing the individual methods to the baseline model trained on the dirty data without cleaning (black dashed line), we can see that there exist certain combinations which result in inferior performance compared to training on the dirty data, which confirms prior studies. 
However, this is not the case for \PaperAcronym{} which always improves the model performance.

An interesting observation in Figure~\ref{fig:exp_data_cleaning} is also that the best data cleaning method also depends strongly on the dataset and type of errors, and thus selecting the best strategy is nontrivial. For instance, the ED2 detector works reasonably well for outliers but results in a deterioration in RMSE for typos (\emph{Airbnb}). In contrast, \PaperAcronym{} finds a competitive combination of detectors and repairs for each data sets; i.e., it chooses different detectors and repairs for each of the datasets, as we can see by studying the weights of the individual methods in Figure~\ref{fig:exp_data_cleaning} (bottom). There is a clear trend that \PaperAcronym{} assigns higher weights to data cleaning methods that perform well in isolation (e.g., DBoost and the missing value detectors and ML imputation strategy for missing values). Interestingly, sometimes the combination (i.e., a mixture) of different data cleaning methods results in a superior overall RMSE than each of the individual methods, as we can see for instance in the case of outliers on the (\emph{Nasa}) dataset.

An interesting future extension of our initial approach could be to also make the actual detection and repair methods also differentiable to train them end-to-end with our approach. A simple example could be to not replace missing values with the mean, but to use a learned value. We believe that there are many more sophisticated approaches to formulate differentiable data cleaning pipelines yet to be explored which can further improve the performance.

\subsection{Learned Dataset Selection}
\label{sec:learned_dataset_selection}

Often, ML engineers need to incorporate training data from multiple sources into model training. Especially if only limited data is available, it often has to be complemented with freely available open data \cite{yakout2012infogather,castelo2021auctus,chepurko2020arda}. 
However, for open datasets the label quality is often not clear and thus additional data might not always improve the downstream ML performance when being included in training \cite{10.14778/2994509.2994514}.
As such, selecting which datasets should be included is another problem that data scientists often need to deal with.\footnote{Note that this problem is also different from data cleaning since low label quality can not be easily addressed with methods to detect and repair data errors, which we discussed in the first use case.}

Traditionally, to decide which datasets should be included in training, we would train the ML pipelines on different combinations of available datasets. However, this is costly since many pipelines need to be trained to identify which subsets of training datasets should be used. 
Hence, in this use case we aim to learn which training datasets should be incorporated, where the challenge is again to formulate the inherently discrete dataset selection step in a differentiable way. 
Our idea is to learn a mixture of models, where each of the models is conceptually trained on a different dataset. 
In the following, we now show how dataset selection can be formulated as a differentiable program. 

\vspace{-1.5ex}\paragraph{Method}
We now introduce our approach to derive differentiable dataset selection. For ease of exposition, we assume that there are only two datasets $D_1$ and $D_2$ and we want to decide which subset of datasets should be used, i.e., either both the datasets or just a single dataset. 
However, our method generalizes to $n$ datasets.

Our idea is based on the following observation: if the subset $D_2$ with erroneous labels is ignored, this is equivalent to not considering the gradients $\nabla_w L(x_i)$ coming from examples $x_i\in D_2$ in the model update phase (i.e., during backpropagation). Hence, by weighting the gradients depending on the dataset they belong to, we can obtain a mixture of models trained on different subsets. 
For instance, if all examples of $D_2$ obtain weight zero, this is equivalent to just training on $D_1$.
More formally, we assign a weight $\lambda_i$ to each dataset and use the following modified update rule instead of standard SGD

\vspace{-4.5ex}
\begin{equation}
\label{eq:update_dataset_selection}
\theta \leftarrow \theta - \eta \frac{1}{n}\sum_{i=1}^{n} \lambda_{d(x_i)} \nabla_\theta L(x_i)
\end{equation}
\vspace{-2.5ex}

where the function $d(x_i)$ returns the index of the dataset the example was sampled from, i.e., $1$ or $2$ for datasets $D_1$ and $D_2,$ respectively. This can be seen as training a mixture of models using the weights $\lambda_{d(x_i)}$ where we weight the gradients based on the source which the data stems from. For instance, if \PaperAcronym{} uses the weights $\lambda_1=0, \lambda_2=1$, we obtain a model that is similar as if it would only be trained on the second dataset whereas for $\lambda_1=0.5, \lambda_2=0.5$ we obtain a model incorporating both datasets equally. Note that a similar weighting of gradients is also compatible with alternative optimization algorithms beyond simple SGD such as Adagrad \cite{DBLP:journals/jmlr/DuchiHS11} or Adam \cite{DBLP:journals/corr/KingmaB14}.
The question now becomes how the weights $\lambda$ can be learned. 

Our idea is to learn the weights $\lambda$ based on how the gradients $\nabla_\theta L(x_i)$ influence the loss of the model. Intuitively, if the model update caused by the gradient of the example $x_i$ increases the loss of the model, the weight of the corresponding dataset should be lowered.
To this end, we first compute the model gradients $\nabla_\theta L(x_i)$ of a data batch as usual, and afterwards draw a second batch for the updated model (according to Equation~\ref{eq:update_dataset_selection}). However, this time we compute the gradient w.r.t. the weights $\lambda$ depending on how the previous $\theta$ gradient updates influence the loss. Accordingly, we update the weights $\lambda$ s.t. gradients from high-quality datasets improving the model performance are preferred.
This method is in particular beneficial if the second batch to update $\lambda$ is drawn from a clean validation set, which could in practice be a small high-quality dataset where the user is sure that it only contains little errors.
As such, we can choose larger weights for datasets, which overall reduces the loss of the model when they are included in the training.
Moreover, similar to the first use case, we do not use the weights directly but the softmax of the weights. 
Overall, the idea of differentiable dataset selection is thus related to methods aiming to assign weights to individual records in the dataset \cite{pmlr-v80-ren18a}. However, our approach is different since we aim to solve dataset selection instead of deciding on a per-example basis whether it should be included.

\begin{figure}
	\centering
	\includegraphics[width=0.8\columnwidth]{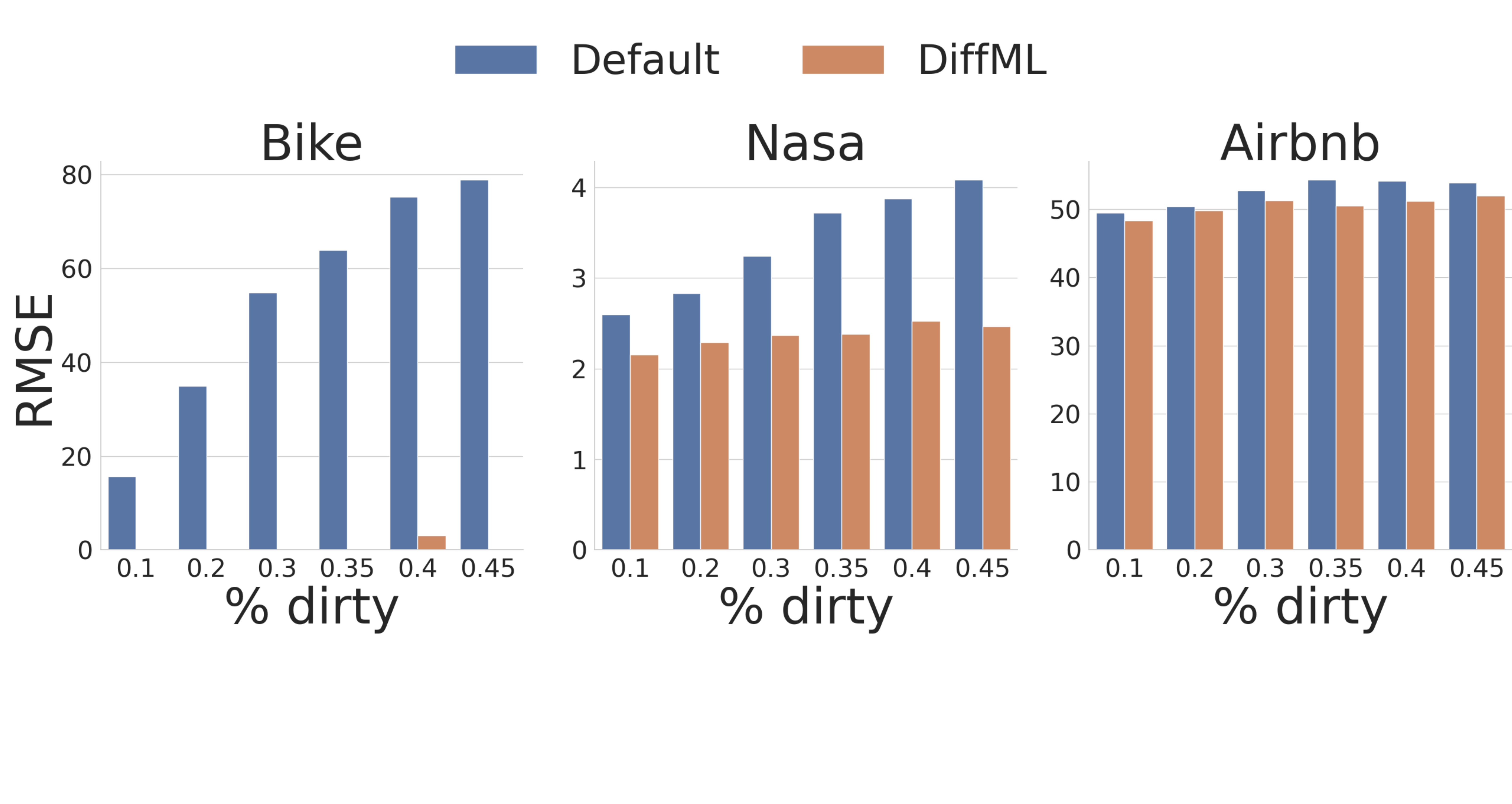}
	\vspace{-8.5ex}
	\caption{Performance of \PaperAcronym for learned dataset selection. In all cases, \PaperAcronym reduces the error (RMSE) by assigning lower weights to the latter one compared to a vanilla model that is trained on a union of the clean and dirty datasets.
}
	\vspace{-4.5ex}
	\label{fig:exp_dataset_reweightning}
\end{figure}

\vspace{-1.5ex}\paragraph{Initial Results.} For the setup in our initial evaluation, we consider two datasets $D_1$ and $D_2$ where the second dataset has low label quality, and it may deteriorate the model performance if it is included in the training. We obtained the datasets by splitting the datasets used in the first experiment in two subsets and perturbing a certain percentage of labels of the second dataset by randomly swapping labels. We then compare the RMSE of \PaperAcronym{}, which can learn to assign lower weights to the dirty dataset, to a baseline method which uses both datasets equally; i.e., the baseline is an MLP trained over the union of both datasets.

As we can see in Figure~\ref{fig:exp_dataset_reweightning}, the proposed method of using \PaperAcronym{} always provides an improvement over the baseline model (referred to as Default in Figure~\ref{fig:exp_dataset_reweightning}) trained on the full dataset. In general, the improvement is dependent on the dataset. For instance, the \emph{Bike} dataset is very sensitive to noise, therefore our method drastically reduces the RMSE. On the other hand, on datasets such as \emph{Airbnb} that are robust to noise (i.e., where the introduced errors do not deteriorate the model performance significantly) the method only yields slightly more accurate results as expected.

\subsection{Learned Feature Selection}
\label{sec:learned_feat_selection}

In ML pipelines, it can sometimes be beneficial to exclude some features from the training process, for instance in cases where features are uninformative or correlated (and thus do not provide additional information). However, selecting the best suited set of features is non-trivial but can have a significant impact on the end-to-end performance of the ML task~\cite{cai2021armnet}. 

State-of-the-art methods for feature selection include filter-based methods, which compute metrics such as correlation with the target variable to select the $k$ best features according to the metric (where $k$ is a hyperparameter), wrapper-based methods \cite{kohavi1997wrappers} which incrementally train models on subsets of features to find a suitable set of features, and embedded methods which exploit characteristics of a model class (e.g., Lasso for linear models, feature importance for decision trees etc.) \cite{hastie2009elements}.
While both filter-based models and wrapper-based techniques require multiple ML pipelines to be trained (which can quickly become costly for a larger number of features), embedded methods are limited to specific model classes and can often not be trained in a differentiable way. An alternative to feature selection are dimensionality reduction techniques such as PCA, which again require multiple pipelines to be trained to find a suitable number of target dimensions $k$.

\vspace{-1.5ex}\paragraph{Method} For this paper, we propose differentiable feature selection, where we again rely on the intuition of mixtures of different ML pipelines. In particular, a feature subset can be seen as setting some features to zero in the feature vectors, which are thus ignored. Hence, instead of using all features in our ML pipeline $\left( x_0, x_1, \dots, x_n \right),$ we again assign a learnable weight to each feature which is normalized between zero and one using the sigmoid function $\sigma(\cdot)$. Hence, we use the feature vector $\left( \sigma(\lambda_0)x_0, \dots, \sigma(\lambda_n)x_n \right)$ in our ML pipeline and again learn the values of the weights $\lambda$ jointly with model training. As such, we can learn that certain features should obtain very low weights (close to zero) while others obtain higher weights.

\begin{figure}
	\centering
	\includegraphics[width=0.8\columnwidth]{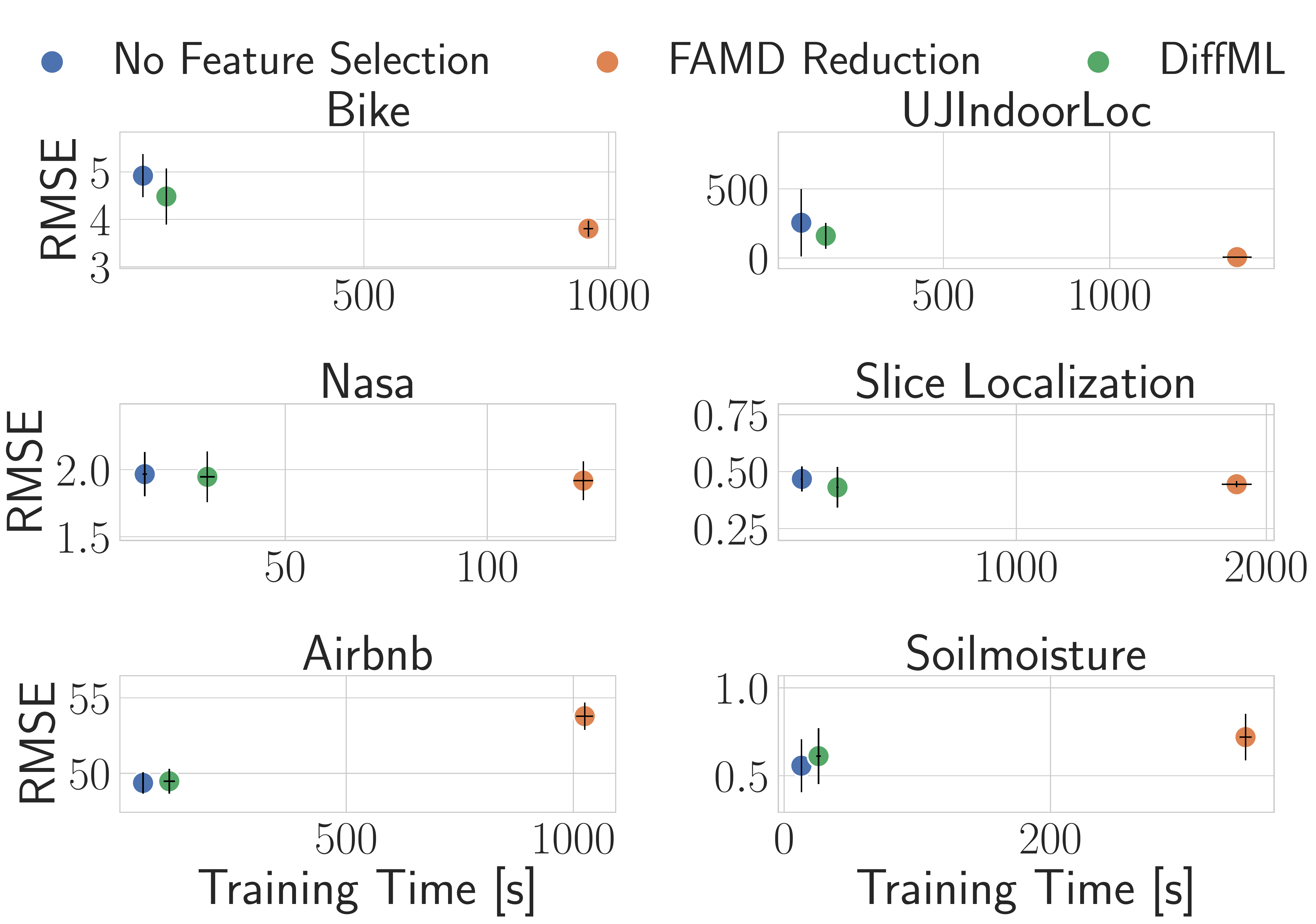}
	\vspace{-3.5ex}
	\caption{Performance and training time of \PaperAcronym for feature selection. \PaperAcronym achieves low errors while only requiring little training time. In contrast, for datasets with many features (right column), FAMD can reduce the prediction errors whereas it can increase the errors in other cases (e.g., \emph{Airbnb}). In addition, it is expensive to train since many pipelines have to be trained. \PaperAcronym{} achieves a competitive performance in both cases while only incurring little overhead to performing no feature selection.}
	\vspace{-4.5ex}
	\label{fig:exp_feature_selection}
\end{figure}

\vspace{-1.5ex}\paragraph{Initial Results.} In our experimental evaluation, we consider the previous datasets as well as three additional datasets: \emph{UJIndoorLoc}, \emph{SoilMoisture} and \emph{Slice Localization} (cf. Figure~\ref{fig:exp_feature_selection}, right column) which have many columns and thus feature selection is more difficult and complex compared to the other datasets. As baselines, we consider no feature selection, as well as FAMD \cite{saporta1990simultaneous}, which is a generalization of PCA to categorical features. We use FAMD to reduce the data to $k$ dimensions (for 15 different values of $k$ and thus 15 resulting ML pipelines which are evaluated). 
While the first baseline (no feature selection) requires only a single pipeline to be trained (like our approach), it does not consider feature selection at all (which can however be beneficial, as we show). In contrast, FAMD requires multiple pipelines to be trained, which can potentially improve the performance but is computationally expensive. We used a timeout of 30 minutes for all techniques. In addition, we implemented forward selection as a baseline, but observed that it was dominated by FAMD in all cases. We thus omitted the results for brevity.

In Figure~\ref{fig:exp_feature_selection}, we present the tradeoff between time spent on feature selection and model training and the resulting accuracy for datasets with few features as well as wide datasets. Overall, our approach achieves competitive predictive performances while incurring almost no overhead compared with the default of implementing no feature selection. In contrast, FAMD is computationally costly since multiple pipelines have to be trained. In terms of accuracy, our approach improves the accuracy of the default case for wide datasets (right column) and has a comparable error for datasets with only few features (left column). In contrast, while FAMD sometimes further improves the accuracy in the case of wide datasets, we can also observe significant performance degradations on datasets with few features when compared to the default of no feature selection. 
Hence, the results for differentiable feature selection are promising since it offers a competitive accuracy improvement in cases where feature selection improves the result, and it does not exhibit regressions in cases with few features and incurs almost no computational overhead.

 \vspace{-1.5ex}\section{The Road Ahead}
\label{sec:res_challenges}

There is still research needed to generalize our findings in the case studies in different directions. The three main research challenges are (i) coverage, i.e., expressing other pipeline steps besides the three of our case studies in a differentiable way, (ii) integration of individual differentiable pipeline steps into full pipelines, and (iii) research beyond the scope of this  vision paper such as other methods beyond mixtures for differentiable ML pipelines. 
In the following, we will describe the open areas in more detail.

\subsection{Other Pipeline Steps}
A first interesting research direction is how to express additional operations beyond the ones presented in our case study such as data augmentation \cite{shorten2019survey,taylor2018improving, miao2021rotom} or data transformation in a differentiable way since this broadens the applicability of differentiable ML pipelines. An initial approach could be to apply our notion of mixtures also to the remaining operators, e.g., by applying our methods for the dataset selection use case on different augmented versions of a dataset to find a suitable set of augmentations. 

While this approach can already extend the set of differentiable pipeline steps, there is a fundamental limitation to a mixtures-based formulation: in a nutshell, we simply learn from a (combination of) alternative pipeline steps which ones to select (e.g., mean imputation vs. KNN imputer for missing values) whereas the alternatives themselves remain static (i.e., either mean or the result of the KNN model or a mix is imputed). An alternative could be to also express not just alternatives of pipeline operators but the operators themselves in a differentiable way (e.g., to use a differentiable model for the imputation) which significantly increases the space of solutions. While this is straightforward to apply for data imputation, it is unclear how to express for instance data augmentation strategies using differentiable models.

\subsection{Complex Multi-Step Pipelines}

In case the pipeline requires more than a single operator for data preprocessing (e.g., data augmentation \textit{and} data cleaning), the question arises how differentiable pipelines can be combined. Before we discuss this question, note that even if just a single pipeline operator is expressed in a differentiable way, the search space of different pipelines is already significantly reduced and this will likely benefit traditional approaches to construct ML pipelines such as AutoML. However, it will likely be much more computationally efficient to express pipelines with multiple steps in a differentiable way and train them jointly with the models. This direction, however, comes with three research challenges we discuss below, namely (i) end-to-end formulation, (ii) theoretical insights, and (iii) efficient inference.

First, it is not straightforward to ensure that differentiable operators are compatible and end-to-end differentiable. For instance, we have to make sure that even though data augmentation will likely have a very different formulation from data cleaning, both can be combined in any order in a full system and still yield a differentiable program. This will require conceptual research, and a first step could be to define a general interface for all differentiable pipeline operators. This area is related to automatic differentiation research \cite{baydin2018automatic,paszke2017automatic}. In addition, theoretical insights are required when the entire optimization is treated as a differentiable program. The reason is that the training will be done with variants of SGD, which is prone to getting ``stuck'' in local minima. While this was shown not be a problem for deep learning \cite{NIPS2016_f2fc9902,DBLP:journals/corr/LuK17}, it is yet to be analyzed for entire ML pipelines.

Finally, there might be optimizations necessary for an efficient inference. For instance, when expressing data cleaning as a mixture and learning which data cleaning method is beneficial, we have to execute all different data cleaning alternatives at runtime and compute the mixture, which can be computationally expensive. An optimization could be to only consider the top-k methods, i.e., with the top-k highest weights. However, depending on how the differentiable formulation is exactly given, more involved techniques might be required.

\subsection{Further Research Opportunities}

We believe that there are many more research opportunities for differentiable ML pipelines on the intersection with different approaches. For instance, the mixture-based formulation, while being general, suffers from the need to explicitly enumerate alternatives that \PaperAcronym{} can choose from. A natural extension to this work can be exploring other strategies to turn the data engineering steps into a differentiable program. Moreover, it is potentially interesting to apply meta learning to differentiable ML pipelines as well, similarly to how meta learning was used for AutoML to leverage previously observed ML pipelines to quickly suggest promising pipelines for new tasks or to bootstrap or prune the search procedure \cite{DBLP:books/sp/19/Vanschoren19}. Finally, in cases where differentiable pipelines should be combined with AutoML (e.g., if not all pipeline operators can be expressed in a differentiable way), research is needed on how to effectively combine the strengths of both approaches. 

\vspace{-1.5ex}\section{Conclusion}
\label{sec:concl}

In this paper, we have presented our vision of differentiable ML pipelines based on the idea of mixtures. This vision has the advantage that the \textit{entire} ML pipeline is optimized for downstream performance, which prevents that pre-processing steps can harm the model accuracy.  In our initial case studies, we have shown how to express data cleaning, dataset selection and feature selection as differentiable programs, which shows the generality of our approach.

We believe that there are many research challenges to enable our vision, both in formulating more pipeline operators in a differentiable way but also supporting complex multistep pipelines. We hope that this research can converge into a new class of systems for differentiable ML pipelines which enable efficient and fully automated construction of ML pipelines with high accuracy and could thus be an important contribution to further democratize ML. 

\begin{acks}
This research is funded by the BMBF project within the “The Future of Value Creation – Research on Production, Services and Work” program. In addition, the research was partly funded by the Hochtief project \emph{AICO} (AI in Construction), the HMWK cluster project \emph{3AI} (The Third Wave of AI), as well as the DFG Collaborative Research Center 1053 (MAKI). Finally, we want to thank hessian.AI at TU Darmstadt as well as DFKI Darmstadt. This work was also supported by the Federal Ministry of Education and Research through grants 02L19C155, 01IS21021A (ITEA project number 20219).
\end{acks}

\balance{}
\bibliographystyle{abbrv}
\bibliography{bibliography}

\end{document}